# Human Activity Recognition for Mobile Robot

Iyiola E. Olatunji
*Department of System Engineering and Engineering Management*
*The Chinese University of Hong Kong*
Shatin, Hong Kong
iyiola@link.cuhk.edu.hk

*Abstract*— **Due to the increasing number of mobile robots including domestic robots for cleaning and maintenance in developed countries, human activity recognition is inevitable for congruent human-robot interaction. Needless to say that this is indeed a challenging task for robots, it is expedient to learn human activities for autonomous mobile robots (AMR) for navigating in an uncontrolled environment without any guidance. Building a correct classifier for complex human action is non-trivial since simple actions can be combined to recognize a complex human activity. In this paper, we trained a model for human activity recognition using convolutional neural network. We trained and validated the model using the Vicon physical action dataset and also tested the model on our generated dataset (VMCUHK). Our experiment shows that our method performs with high accuracy, human activity recognition task both on the Vicon physical action dataset and VMCUHK dataset.**

## I. Introduction

Human activity recognition (HAR) is important for autonomous robot's interaction of with real world object, environment and people to further enhance its capabilities. Human activity recognition involves the interpretation of human actions or gesture from a series of human activities. Daily human tasks can be simplified and automated if they are correctly recognized. Complex human activities can be subdivided into simpler ones which can later be combined to solve complex activity recognition problem [1]. For example, a domestic service robot for cleaning can recognize a sequence of activity as "sitting" followed by "standing" and "walking" (meaning the person has left the current environment) and discretionally clean up as a response to such combination of activities. This has several applications in catering for needs of elderly people living alone or people with disability.

3D human action classification problems have been studied using various methods. Bayesian classifier has been used on multiple calibrated cameras for human gesture classification using motion history and energy volumes [1]. In [2] each gesture is decomposed into atoms and Hidden Markov Model (HMM) was used to identify their temporal evolution. Multi-Class AdaBoost algorithm was used in [3] together with dynamic programming algorithm for 3D joint feature processing and increased recognition accuracy. [4] compared 3 dynamic neural networks; Focused Time-Delay Neural Network (FTDNN), Layer-Recurrent Neural Network (LRNN) and Distributed Time-Delay Neural Network (DTDNN) for classification of human activities. Generic programming has also been used for human action recognition [5]. In this paper, we trained a convolutional neural network (CNN) to correctly classify human activities for recognition task.

The problem can be defined as: given an input of 3D human action sequence, the classifier should identify the sequence of activities performed from the kinematic data of human actions.

The objective is to correctly classify 3D human activities from kinematic data. As the first step to recognizing sequence of activities, a model for individual activity recognition task was first built. The goal is to combine individual human action for activity recognition for mobile robot.

The rest of this paper is organized as follows: Section II describes the model and parameters used; Section III describes the experimental set up, data and data preprocessing task; Section IV discusses the findings; finally, Section V concludes the paper.

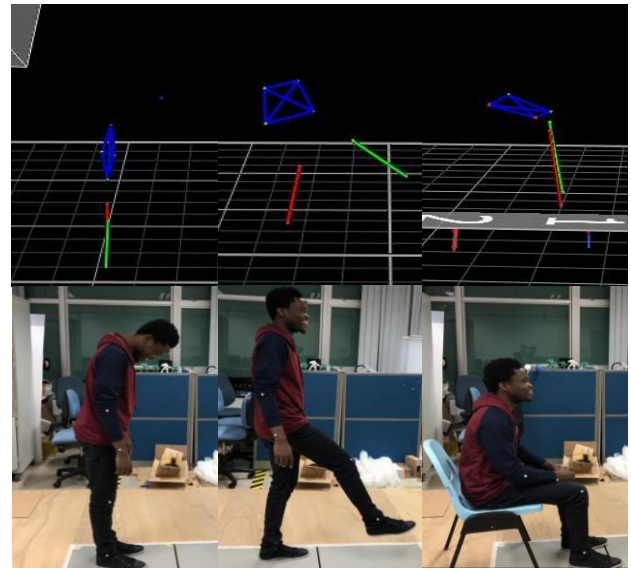

Fig. 1. Action representation on Vicon system and ground truth (VMCUHK)

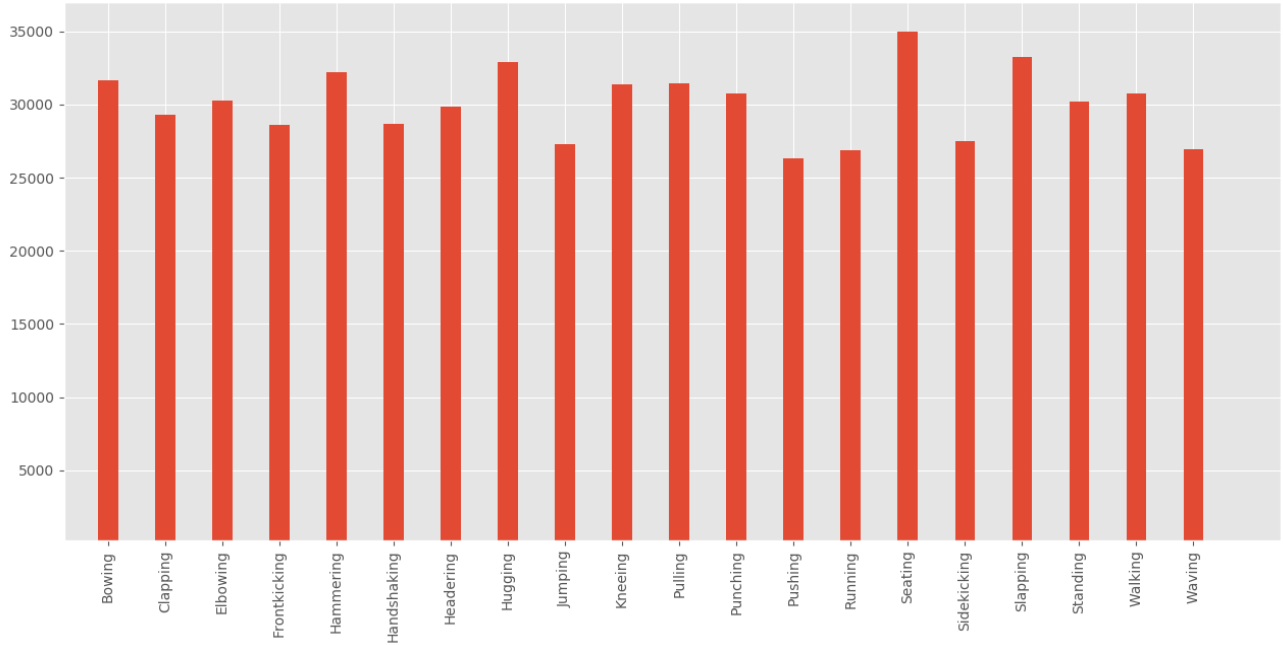

Fig. 2. Data distribution of different activities on Vicon physical action dataset

## II. MODEL DESCRIPTION AND PARAMETER SETTINGS

Deep learning models are a class of machine learning models that can learn features from raw data. Deep learning techniques have been applied to several domains such as object recognition [6], human tracking [7], natural language processing [8], human action recognition [6] and has shown competitive performance.

Convolutional neural networks (CNNs) are a type of deep models that uses multiple layers, shared weights, local connections in the feature maps (units) of the previous layer through filter bank (set of weights) for learning complex features and a pooling layer to alternatingly combine semantically similar features into one from raw input [9].

The features used in the training of our models are the kinematic data of the head, arms and legs. These served as input to the neural network. The architecture of the model consists of one dimensional 2 convolutional layers, 1 max pooling layer and 1 fully connected layer connected to the Softmax layer.

We used depthwise convolution which is an effective method to reduce computational complexity of deep neural networks. It consists of independently performed spatial convolution over each input channel followed by a 1x1 convolution output channel [10]. This output serves as input to the Rectified Linear Unit (ReLU) activation function and we performed 1D max pooling on the output of convolution layer. The filter size and depth of the first convolution layer is set to 60 and the filter size of the pooling layer is 20 with a stride of 2. The second convolutional layer has a stride of 6 and the output is flattened out for the fully connected layer input. The fully connected layer consist of 1000 neurons and tanh function is used for non-linearity. The Softmax layer produces probabilities of class labels.

Stochastic gradient descent is used for minimizing the cost function. The model is trained using a batch size of 200 and 1000 epochs.

## III. DATA PREPROCESSING AND EXPERIMENTAL SETUP

Our activity recognition model was trained and tested on the publicly available Vicon physical action dataset [11] used for mobile robot. We also carried out our own experiment to validate the accuracy of the trained model by generating 6 different activity data using the Vicon motion capture system from the computer aided design laboratory of the Mechanical and Automation Engineering (MAE) department of the Chinese university of Hong Kong (VMCUHK). Our result shows that the trained model can recognize similar activities with some degree of variation with high accuracy.

### A. Model Validation through Vicon physical activity dataset

The Vicon physical action dataset [11] consists of 10 normal and 10 aggressive physical actions that models human activity collected from 10 subjects (7 males and 3 females) using the Vicon Motion Capture System. Twenty distinct experiments were performed by each subject which are grouped into 10 normal and 10 aggressive activities and the experimental trial were separately taken for each physical activity. Each activity lasted for approximately 10 seconds per subject at a sampling frequency of 200Hz which amounted to approximately 3000 samples per subject in an activity. An average of 15 normal and 15 aggressive actions trajectories were extracted within the 10 seconds of each activity. Twenty-three markers were used to create 6 kinematic models, 2 × lower legs, 2 × lower arms, 1 × head and 1 × shoulder for generation of human activity data.

In our experiment, we took a different approach by classifying into 20 classes instead of the 2 classes of aggressive or normal activity of [11]. We exploited all activities and trained our model on 20 classes of activities namely; Bowing, Clapping, Elbowing, Front kicking, Hammering, Handshaking, Heading, Hugging, Jumping, Kneeing, Pulling, Punching, Pushing, Running, Seating, Side kicking, Slapping, Standing, Walking, and Waving. This was done to fully model a broad range of human activity for better simplified action recognition for autonomous mobile robots which can be combined to form a sequence of activity.

*B. Model Testing through VMCUHK*

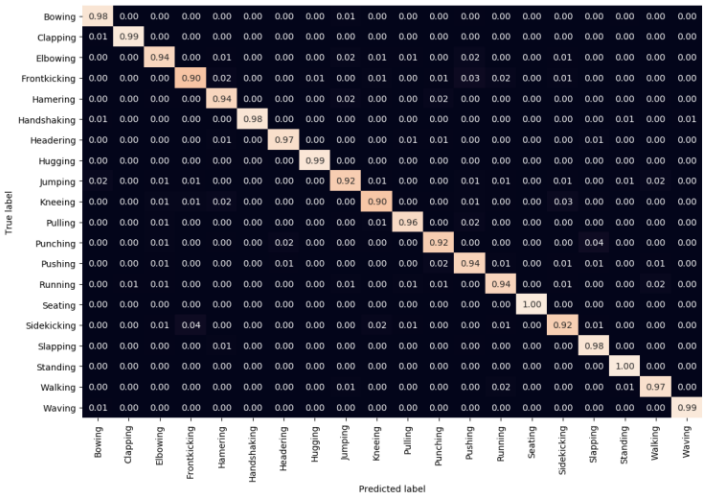

Fig. 3a. Confusion Matrix of Vicon physical activity dataset

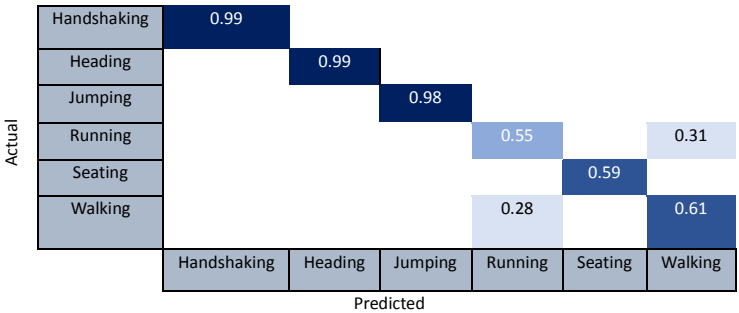

Fig. 3b. Confusion Matrix of VMCUHK

Vicon Motion Capture System from the computer aided design laboratory of the Mechanical and Automation Engineering (MAE) department of the Chinese university of Hong Kong (CUHK) was used for the generation of our dataset similar to [5] but our setup is relatively different. We used 8 Vicon cameras instead of 9 cameras used in [5] for the generation of the kinematic data. In [5], 23 markers were used to create kinematic model but we used 9 markers to create our kinematic model. The markers were placed on the elbow and wrist of the arms and knee and ankle of the legs as shown in figure 1. VMCUHK only consist of 6 activities namely: handshaking, heading, jumping, running, seating and walking.

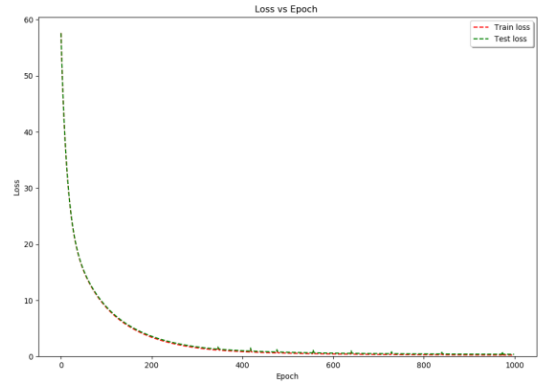

Fig. 4a. Comparison of training loss with test loss

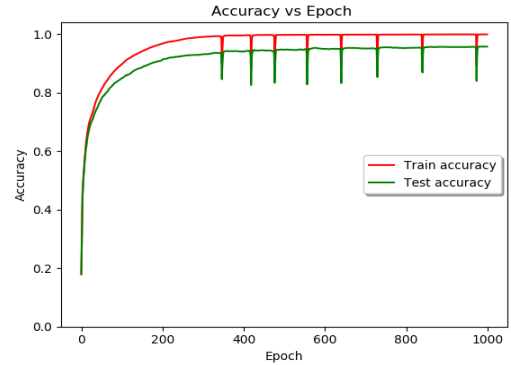

Fig. 4b. Training accuracy comparison with test accuracy

## IV. EXPERIMENTAL RESULT AND ANALYSIS

Figure 1 shows the action representation of 3 physical actions expressed by 3D kinematic human models and the ground truth underneath on VMCUHK dataset.

We evaluated the soundness of our model by generating kinematic data of previously unseen activities (VMCUHK dataset). Random splitting of the dataset with 70% for training and 30% for testing was performed on the Vicon physical action dataset. The number of labels (classes) is 20 and learning rate was set to 0.0001. We trained the model on Nvidia GeForce GTX 1080 GPU using TensorFlow library [12] and used Adam Optimizer for optimizing convergence. We used L2 loss and softmax cross entropy with logit function from the TensorFlow library to measure the error probability of mutually exclusive classes. One hot encoding was used to better prediction of classes. The training time was about 1 hour. However, the model converged at about 600 epochs but it was left to run for 1000 epochs thus the training time could be significantly reduced to half if early stopped.

As shown in figure 3a, our model can correctly classify the unseen activity data from Vicon physical action dataset with an accuracy of 95% and high precision as shown in Table 1.

Figure 3b shows the confusion matrix of testing our model on VMCUHK. This shows that our model can recognize similar activities with some degree of variation considering the huge disparity between our setup and Vicon physical action data generation. Misclassification on VMCUHK is mainly on very similar actions. For example, the action "walking" and "running". Also "seating" and "kneeing but the kneeing action was not performed on VMCUHK.

| Accuracy | 95.71% |
|---|---|
| Precision | 95.72% |
| Recall | 95.71% |
| F1 score | 95.70% |

Table 1. Experimental result on Vicon physical action dataset

## V. CONCLUSION AND FUTURE WORK

The efficiency of domestic service robots can be improved if they can easily recognize complex human actions and thus foster harmonious human-robot interaction. Needless to say that this is indeed a challenging task for robots, it is expedient to learn human activities for autonomous mobile robots (AMR) for navigating in an uncontrolled environment without any guidance.

We trained a model using depthwise convolution neural network for human activity recognition. Result shows that our method performs with high accuracy, human activity recognition task both on the Vicon physical action dataset and VMCUHK dataset and similar activities with some degree of variation can be recognized with high accuracy.

Expanding the VMCUHK dataset to all classes instead of the current 6 classes will be considered in a future work.

*A. Acknowlegement*

Special thanks to Fengyan Liang for his technical support in conducting experiment with the Vicon system.